\documentclass[conference]{IEEEtran}
\IEEEoverridecommandlockouts
\usepackage{cite}
\usepackage{url}
\usepackage{amsmath,amssymb,amsfonts}
\usepackage{algorithmic}
\usepackage{graphicx}
\usepackage{textcomp}
\usepackage{xcolor}
\def\BibTeX{{\rm B\kern-.05em{\sc i\kern-.025em b}\kern-.08em
    T\kern-.1667em\lower.7ex\hbox{E}\kern-.125emX}}

\begin{document}

\title{D\'ej\`a-Vu: A Glimpse on Radioactive Soft-Error Consequences on Classical and Quantum Computations}

\author{Antonio Nappa, Christopher Hobbs, Andrea Lanzi}

\maketitle
\thispagestyle{plain}
\pagestyle{plain}
\begin{abstract}
What do Apple, the FBI and a Belgian politician have in common?
In 2003, in Belgium there was an election using electronic voting machines. Mysteriously one candidate summed an excess of 4096 votes. An accurate analysis led to the official explanation that a spontaneous creation of a bit in position 13 of the memory of the computer attributed 4096 extra votes to one candidate. One of the most credited answers to this event is attributed to cosmic rays i.e.(gamma), which can filter through the atmosphere. Indeed, SEUs (single event upsets) such as bit-flips are frequent faults especially in outer space, where ionizing radiations are very frequent.
For this reason many space capable devices mount very expensive and technically advanced memories, to reduce risks of malfunction, breakage, or exploit.
On the other hand, at ground level, consumer devices aren't equipped with such memories for economic and statistical reasons. Indeed, research in the area shows that cosmic, high-energy rays filter inside the atmosphere on a daily basis. But when they hit our computers mostly they reach non-allocated space and systems do not notice their effects.
There are cases though, with classical computers, like forensic investigations, or system recovery where such soft-errors may be helpful to gain root privileges and recover data. In this paper we show preliminary results of using radioactive sources as a mean to generate bit-flips and exploit classical electronic computation devices. \textit{We used low radioactive emissions generated by Cobalt and Cesium and obtained bit-flips which made the program under attack crash.} We also provide the first overview of the consequences of SEUs in quantum computers which are today used in production for protein folding optimization, showing potential impactful consequences.
To the best of our knowledge we are the first to leverage SEUs for exploitation purposes which could be of great impact on classical and quantum computers.
\end{abstract}


\section{Introduction}
In 2016 the FBI demanded that Apple unlock (decrypt) an iPhone 5C after the San Bernardino's terrorist shooting~\cite{fbi}. Apple, denied the decryption of the phone alleging that it would put the privacy of millions of users at risk. According to the news, the FBI resorted to an Israeli company, which charged around 1M USD. The company reverse engineered the  phone, which allowed the investigation to continue.
This study strives to leverage radioactive energy emissions to obtain bit-flips on electronic devices e.g.(mobile phones, electronic voting machines, IoT devices) for exploitation without resorting to expensive and ad-hoc techniques. Also it shows how such radioactive emission can impact quantum computation in a significant and dangerous way.
Our preliminary research shows that on memories (+15 years old) low radioactive emissions (within 0.4MeV and 1.4 MeV of kinetic energy) produce SEUs and lead a program to a crash, while on modern memory such energy is not sufficient.

Soft-Errors are a type of hardware fault that in some cases provoke a glitch\footnote{malfunction} in a system executing a program. A glitch normally can manifest as a bit flipping its value from 0 to 1 or vice-versa. The intuition of soft-errors generated by radioactivity is attributed to the Sci-Fi writer Dr. Isaac Asimov, which in his book "Caves of Steel"~\cite{caves} (written in 1953), reports hardware malfunctions due to radioactivity. Scientific proofs of such evidence came years later. First by observing how impurities in construction of silicon wafers could autonomously generate $\alpha$ particles\footnote{An $\alpha$ particle corresponds to a 2 proton 2 neutron particle, similar to $^{4}He$, the most common form of Helium on Earth} which could lead to soft-errors~\cite{seminal, seminal2}, these particles are not dangerous for human health and carry very low energy, measured in MeV. As a matter of fact $\alpha$ particles  can be shielded with a sheet of paper.
Along the past century the advent of space missions made the phenomena easier to observe at high altitudes and outside Earth's atmosphere. Indeed, in outer space high energy cosmic rays are free to hit unshielded electronic circuits and provoke glitches, errors, and faults. Recently, the introduction of quantum computers opens new avenues for these source of errors to generate potential harmful issues.
In this paper we report preliminary results of inducing soft-errors by hitting classical circuits with gamma-rays and low energy neutrons to break into protected systems, leveraging bit-flips and following exploit strategies similar to RowHammer~\cite{rowh}. Our technique can be leveraged for forensic investigation on locked/encrypted electronic devices, to recover data, or to rescue a system. Moreover we provide an overview on how such technique could impact quantum computations.

Our hypothesis is to prove that high power emissions (available at radiotherapy clinics at an affordable cost~\cite{cheapseu, clinical}) can be used to exploit COTS (Common Off The Shelf) electronic devices and consequently allow their exploitation~\cite{rowh, rowh2, drammer, drammer} or lead to wrong output of quantum computations. This is the first attempt in scientific literature to break into a device i.e.(classical or quantum) using radioactive sources.


\section{Related Work}
Literature related to Soft-Errors appeared first in the 70s~\cite{seminal2,seminal,retro}, where such errors were generated by impurities in the manufacturing package of memory and CPU chips. In this specific case the SEU was induced by an $\alpha$ particle. Over time, hardware manufacturers have been improving the selection of their materials and these kinds of soft errors have disappeared. Nonetheless, in a paper from 2017~\cite{cheapseu}, the authors show that by removing the epoxy layer of a chip, makes very easy to generate soft errors with $\alpha$ particles. Such particles are quite common in nature even at ground level, and hence removing the epoxy leaves the chip exposed even to innocuous radiations. Along with research for hardware that would function at ground level there is a huge corpus of research related to hardware that needs to function from the stratosphere to outer space. Indeed, many papers~\cite{outer1,outer2,outer3} especially in the field of particle physics and radiation protection for electronic devices, have shown that it is quite common for such errors to appear. Research efforts have been focused on observing/producing such errors at ground level with the help of radioactive sources. In a paper from 2005 of Franco et al.~\cite{clinical} show how to generate soft errors in commercial SRAM with low energy neutrons produced at a clinical facility. With respect to SDRAM which is the most common memory used in COTS devices, a 2018 paper from Kohler et.al \cite{ddr} shows that multiple kinds of errors (e.g. single-event effects (SEE), single bit upsets (SEUs), multiple-bit upsets (MBUs), multiple cell upsets (MCUs), latch-up (SEL), and functional interrupt (SEFI).) can be reproduced by using ionizing radiations emitted by devices in a clinical facility treating oncological patients. Schroeder et al.~\cite{toronto} show how the impact of cosmic rays on electronic devices change between sea level and above, closer to space. On a daily basis, computers are hit by cosmic rays which can provoke glitches, but in most cases they hit memory which is free, so the running system is not impacted. Indeed, a paper from 2011 shows how such phenomena can influence the DNS system~\cite{bitsqua}. Tezzaron Semiconductor in 2004 reports in a white paper how such errors may appear within a a range of 0.2 and 1 per day in COTS memories~\cite{tezzaron}. Even if there is no evidence of effects of SEUs in quantum hardware in literature it is evident that everything that is based on silicon and similar semiconductors is subject to the same physical laws.


\section{Approach}
The intuition of this work, is to leverage ionizing radiations to induce bit-flips in specific areas of classical computer memory. We have devised two techniques for this attack to take place. One is by using page table spraying~\cite{rowh} and the second requires minimal reverse engineering of the memory locations of the DUT (Device Under Test). The first attack works where there is at least user-level privilege, while the second works when the device is locked. We have performed multiple tests with low energy ionizing radiation (within 0.4 and 1.4 MeV) with Cesium and Cobalt on the Raspberry Pi 4 platform with 8 GB and 4 GB of memory. We also have tested an older IBM ThinkPad T41p, which was manufactured in 2003. Our final goal beyond the proof of concept of exploiting a user-land program similar to the RowHammer~\cite{rowh} technique, is to unlock a mobile phone or any kind of device which is protected and needs to be unlocked for legal reasons or for data rescue. Our technique also impacts quantum hardware given its high sensitivity to environmental variations, see Section~\ref{sec:quantum} for detailed explanation.

\begin{figure}[h]
\includegraphics[width=0.45\textwidth]{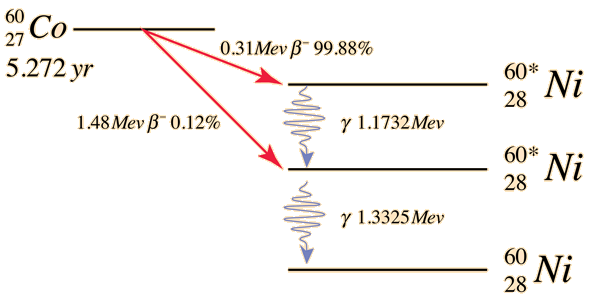}
\caption{Cobalt Decay Ratio}
\label{fig:ratio}
\end{figure}

To explain the nature of radioactive ionizing emissions, Figure~\ref{fig:ratio} shows the decay of Cobalt, which emits beta and gamma rays. Radioactive materials are unstable by nature and they evolve into other materials over time by losing energy in the form of particles (i.e. electrons). Hence, by losing electrons one material transforms into another over time. For example Cobalt-60 isotope after approximately 5 years transforms into Nickel-60. Figure~\ref{fig:ratio} shows the decay of Cobalt which produces different source of energy, $\beta$ and $\gamma$ radiations, which carry different kinetic energies, between 0.31 MeV and 1.33 MeV. Radioactive decay produces ionizing radiations which affect electronic circuits as shown in many papers~\cite{cazza, cazza1, cheapseu, clinical} and in circuit manufacturing recommendations guides~\cite{itu}.

The ratio of these emissions varies according to the materials. The more unstable and dangerous the materials are, the more radiations emits. Enhancing the probability of successful SEUs generation. Understanding the ratio of our soft-error inducing approach, requires time and study of materials and different devices along with their manufacturing process. This holds true in both scenarios i.e. (classical and quantum computers).

\begin{figure}[h]
\includegraphics[width=0.5\textwidth]{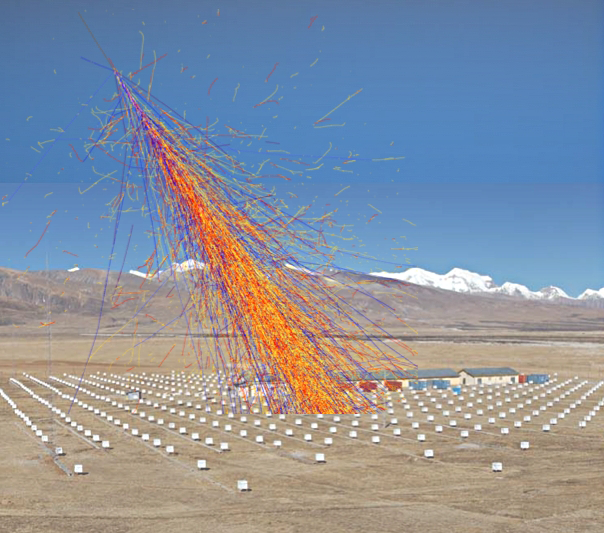}
\caption{High Energy Experiment Realized in Tibet in 2019}
\label{fig:tibet}
\end{figure}

\begin{figure}[h]
\includegraphics[width=0.45\textwidth]{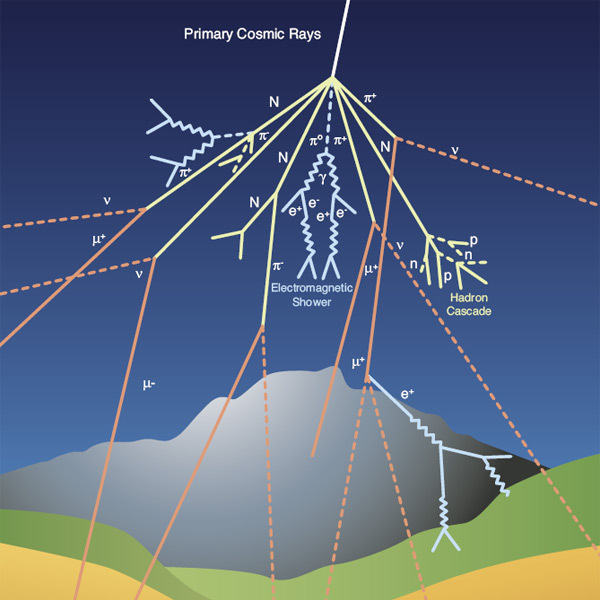}
\caption{Example of Cosmic Ray Decay in Earth's Atmosphere}
\label{fig:atmo}
\end{figure}

\paragraph{\bf Bit-flips in the wild.}
Multiple papers~\cite{ground, tezzaron} show that SEU happen spontaneously due to cosmic radiations, coming from sources in outer space. More recent papers~\cite{cazza,cazza1}, show that errors can be induced even in very recent electronics. In another paper~\cite{tibet}, shown in Figure~\ref{fig:tibet} it was possible to observe in the geographical zone of Tibet, radiations beyond 100TeV of energy which could eventually permanently damage the circuits. However, the number of photons with such energy was 24, which still is very small compared to the number of bits that are present in COTS memory chips. Figure~\ref{fig:atmo} shows how cosmic rays when hitting our atmosphere generate different sources like a particle accelerator does when particles collide. Daily, some of these secondary charges impact classical computer memories and generate SEUs, but the low probability of such events to happen at a sufficient rate makes natural sources not reliable to be used as a potential exploit avenue, at least in classical computing.

\paragraph{\bf Bit-flips from radioactive sources.}
These events are intentionally generated with the help of radioactive materials, at a very small distance from the DUT in a controlled laboratory and very often in vacuum, either with strong radiation sources~\cite{cazza, cazza1, clinical} or with innocuous $\beta$ radiations by removing the epoxy layer of a chip~\cite{cheapseu}. Some of these experiments demonstrate the ability of generating failures in ML models and opening space for non conventional adversarial attacks. In 2020 Oliveira et al~\cite{cazza}, show how both thermal neutrons and high energy neutrons can generate different kinds of soft errors and alter the output of a computation. The ITU (International Telecommunication Union) recommends~\cite{itu} to test electronic devices against ionizing radiations with a particle accelerator with different sources and secondary charges generated by hitting other particles such the ones contained in water (Oxygen and Hydrogen). Figure~\ref{fig:dut} shows our experiment with Cobalt and water to generate a secondary source thermal neutrons. Table~\ref{tab:itures} shows the results of the ITU test with different kinds of silicon hit by a neutron (+n) carrying different energies generated from a particle accelerator such as the one found in hospitals for cancer treatment. Such experiments result in an emission of an atom of a different material such as Magnesium (Mg) or Aluminium (Al) and a proton (p) or an alpha particle ($\alpha$) and possible generation of SEUs.

\paragraph{\bf Quantum Computation Impact.}
The consequences of radioactive emissions, induced on purpose or generated by natural events on quantum hardware have not been studied and understood yet. Though, the impact of errors and their correction in quantum computations play a fundamental role~\cite{qcorr} in such machines. In Section~\ref{sec:quantum} we give a first intuition on the potential impactful consequences of quantum circuit malfunctioning.

\begin{table}[h]
\centering
\begin{tabular}{|l|l|}
\hline
\multicolumn{1}{|c|}{\textbf{Si - Neutron reactions}} & \multicolumn{1}{c|}{\textbf{Energy}} \\ \hline
28Si + n --\textgreater 28Al + p                                              & 3.999 MeV                                               \\ \hline
28Si + n --\textgreater 25Mg + $\alpha$                                       & 2.749 MeV                              \\ \hline
29Si + n --\textgreater 29Al + p                                              & 3.009 MeV                              \\ \hline
29Si + n --\textgreater 26Mg + $\alpha$                                       & 35.00 KeV                              \\ \hline
30Si + n --\textgreater 30Al + p                                              & 8.040 MeV                              \\ \hline
30Si + n --\textgreater 27Mg + $\alpha$                                       & 4.341 MeV                              \\ \hline
\end{tabular}
\caption{ITU Results of bombarding different Silicon Isotopes with neutrons at different energies}
\label{tab:itures}
\end{table}

\subsection{Soft-Error Hunting on Classical Computers}
Understanding the precise location of a single bit-flip is quite difficult given the density of recent memory chips and their sizes. However, previous approaches~\cite{drammer} have found avenues to reproduce bit-flips in a deterministic way leveraging hardware manufacturing imperfections. In our case, which uses external radioactive sources, the position of a single bit may be smaller than a square nanometer. Researchers~\cite{ddr,laser} use lasers and Monte Carlo simulation to understand the precise location of an error. To this extent, Carmon et al.~\cite{fotonico} show how to recover memory content with photonic side channels, which can help to understand precise bit locations on a small budget. For our preliminary experiment we chose a program to read memory and compare the initial pattern, for example 0xFF and checking its consistency during the irradiation period. Previous literature~\cite{cheapseu} shows that different initial patterns may lead to different error rates. The first RowHammer~\cite{rowh} implementation increases the attack likelihood from 2\% to 30\% by spraying the page table structure of a benign program in user space. Creating a reference to the program memory in a large section of the available free memory, which is bigger than the memory occupied by the program itself. By spraying the page table, it is possible to hit a sensitive bit once or many times which can be flipped and used to escalate privileges. Other approaches~\cite{laser} use high precision laser which could lead to deterministic bit-flips and reduce risks due to radioactivity.
Our initial approach uses ionizing i.e. (emitting particles) radiations which exhibit a statistical behavior as the quantum mechanics theory describes~\cite{pqm}. Our intuition is to shield eventual sensitive regions of the circuit board with lead and and hit only the memory chip. In this way, we are very likely to accomplish exploitation without altering the correct duty-cycle of the DUT. As demonstrated previously~\cite{rowh,rowh2} by spraying the page table of a \textit{suid} program through 4GB of memory gives a 30\% chance to hit the correct byte and escalate privileges.

\subsection{Exploit Strategies on Classical Computers}

\begin{figure}[h!]
\includegraphics[width=0.45\textwidth]{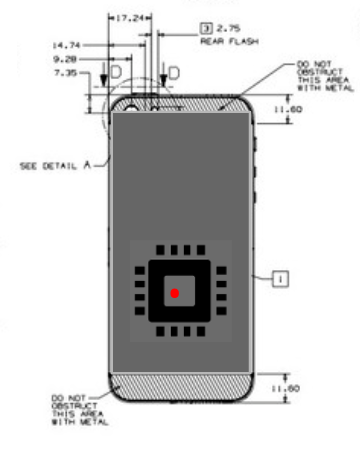}
\caption{A Sketch on how increase probability to shoot on target using lead shields}
\label{fig:shield}
\end{figure}

\paragraph{\bf Basic Reverse Engineering + Bit-Flip.}
This novel technique requires basic knowledge of the DUT, particularly its architecture and some basic information that can be find in the firmware. Indeed firmware images~\cite{amat,gent} carry very specific information within the binary file. For example in baseband chips such as Samsung's Shannon (ARM Cortex R7) there's no page table and MMU (Memory Management Unit). Hence, some basic knowledge of the firmware and the architecture can lead to a successful exploit starting from a SEU. The difficulty is to understand where exactly it's necessary to produce the SEU. For this reason we have proposed shielding the device with lead as shown in the sketch of Figure~\ref{fig:shield}.

\paragraph{\bf Bit-Flip + Privilege Escalation.}
This technique has been already demonstrated by Seaborn et al.~\cite{zero}. In our case, we leverage a different source of SEUs (i.e. instead of RowHammer~\cite{rowh} we use ionizing radiations). The feasibility of our approach resides in the ratio of SEU generation with respect to the size of page-table spraying.

\section{Preliminary Evaluation on Classical Hardware}
\label{sec:eval}
We have used three different devices:  an IBM Thinkpad T41p of 2003 equipped with 2GB of RAM, a more recent Raspberry Pi 4 with 4GB of RAM, and a Raspberry Pi 4b with 8GB of Ram. 

\begin{figure}[h!]
\includegraphics[width=0.45\textwidth]{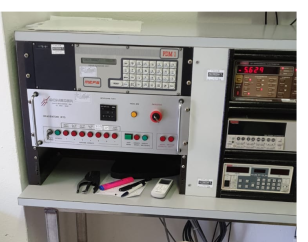}
\caption{Emission Control Console}
\label{fig:console}
\end{figure}

\begin{figure}[h]
\centering
\includegraphics[width=0.45\textwidth]{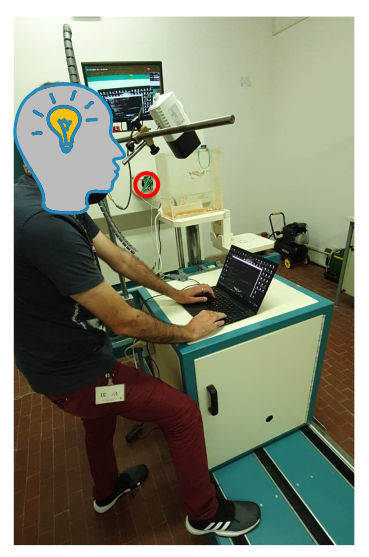}
\caption{Our Experiment testbed, DUT circled close to the radioactive sources}
\label{fig:experiment}
\end{figure}

Tables~\ref{tab:t41co} and ~\ref{tab:t41cs} show the results of gamma-ray exposure of the T41p with Cobalt and Cesium respectively. As a first experiment we filled the memory with different patterns 0xFF, 0X00, 0X41 which generate different bit patterns. Every pattern was tested over a period of 20 minutes with a Cobalt and Cesium source. We were able to observe a total of 68 bit-flips with Cobalt and 13 with Cesium.
After the preliminary successful test the device was ready to be exploitable with the technique explained by Seaborn et al.~\cite{zero}, our attempt with gamma-rays \textbf{lead the \textit{ping} program (which runs with root privileges) to crash two times with segmentation fault}.
The probability of exploitation given the low number of generated SEUs per second (1 every 17 seconds on average), combined with RAM refresh rate (64ms), page table size and RAM constructive process, gave us approximately an average probability of 1\% to shoot on target. More sophisticated techniques, such as lasers~\cite{laser, ddr} may lead to high probability or even deterministic success rate.
It is also worth noting, as it is possible to evince from Figure~\ref{fig:experiment} that all the SEUs were observed with the DUT as close as possible to the to the radioactive source. Figure~\ref{fig:experiment} shows a DUT at less than 5cm from the radioactive source. The laboratory was equipped with a camera and a secondary laptop connected via Ethernet cable to the DUT. The camera was pointing at the screen of the laptop connected to the DUT to observe in real time the result of the experiments. Moreover Figure~\ref{fig:console} shows the console to select duration and radioactive source for the experiment.

\begin{table}[h!]
\centering
\begin{tabular}{|l|l|c|l|l|c|}
\hline
\textbf{Element} & \textbf{Device} & \textbf{Used Memory} & \textbf{Pattern} & \textbf{Time} & \textbf{Flips} \\ \hline
\begin{tabular}[c]{@{}l@{}}Co($\sim$1.5MeV)\end{tabular} & T41p & 1GB  & 0xFF   & 1200s         & 27             \\ \hline
\begin{tabular}[c]{@{}l@{}}Co($\sim$1.5MeV)\end{tabular} & T41p & 1GB  & 0x00   & 1200s         & 22             \\ \hline
\begin{tabular}[c]{@{}l@{}}Co($\sim$1.5MeV)\end{tabular} & T41p & 1GB  & 0x41   & 1200s         & 19             \\ \hline
\end{tabular}
\caption{IBM T41p - Cobalt Gamma Ray Exposure}
\label{tab:t41co}
\end{table}

\begin{table}[h!]
\centering
\begin{tabular}{|l|l|c|c|l|c|}
\hline
\textbf{Element} & \textbf{Device} & \textbf{Used Memory} & \textbf{Pattern} & \textbf{Time} & \textbf{Flips} \\ \hline
\begin{tabular}[c]{@{}l@{}}Cs($\sim$0.6MeV)\end{tabular} & T41p & 1GB  & 0xFF   & 1200s         & 7             \\ \hline
\begin{tabular}[c]{@{}l@{}}Cs($\sim$0.6MeV)\end{tabular} & T41p & 1GB  & 0x00   & 1200s         & 3             \\ \hline
\begin{tabular}[c]{@{}l@{}}Cs($\sim$0.6MeV)\end{tabular} & T41p & 1GB  & 0x41   & 1200s         & 3             \\ \hline
\end{tabular}
\caption{IBM T41p - Cesium Gamma Ray Exposure}
\label{tab:t41cs}
\end{table}

\begin{figure}[h]
\centering
\includegraphics[width=0.40\textwidth]{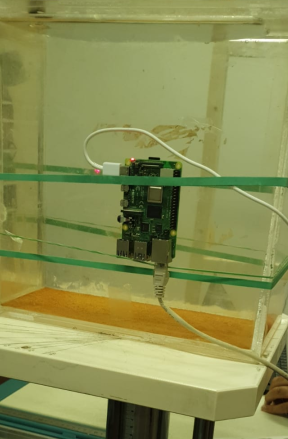}
\caption{Raspberry Pi4B Under Test with a Water Tank}
\label{fig:dut}
\end{figure}

The experiments with the more recent (dated 2019) and more dense memories (higher bitrate for square millimeter) of the Raspberry Pi devices led to zero SEUs. According to the ITU report~\cite{itu} from 2018 higher energies (above 5MeV) and heavier particles are needed i.e. (neutrons instead of photons) to generate SEUs. Nonetheless even using a water tank as shown in Figure~\ref{fig:dut} was not sufficient to generate secondary charges which could significantly impact the memory of the DUT.
Moreover we left one Raspberry Pi 4b with 8GB of memory with 4GB allocated with the pattern 0xFF during 10 months without noticing any SEU. This does not exactly means such events did not happen, but reading the entire memory takes time and it is a sequential operation, per byte operation. Hence our reading single threaded loop routine may just have missed the SEUs because of its delay in reading and comparing memory values.

\section{Preliminary Discussion and Future Work}
\subsection{Classical Computers}
Our preliminary experiments show that our approach it is likely to work (81 bit-flips were generated with gamma-rays) on 20 years old devices and we successfully crashed a \textit{suid} program 2 times. But for recent devices it will be necessary to access heavier particles i.e. (neutrons or protons) instead of photons (gamma-rays) and higher kinetic energy (above 5MeV). The cost of such a technique is much cheaper than performing a complete reverse engineering of the device memory looking for a vulnerability. A few sessions of advanced cancer treatment (hardron-therapy)~\cite{hardron} cost about a tenth ($\sim$15K USD) of what any reverse engineering company may ask from law enforcement to unlock a device ($\sim$1M USD)~\cite{fbi}. Moreover our technique is extremely portable, fast and it needs small customization. It requires knowledge of the memory structure and architecture of the DUT. Such information is stored directly in firmware images while other information maybe inferred from similar unlocked devices.
We contacted multiple clinical facilities across Europe asking to test our approach with hardron-therapy. In Europe there are only about fifteen centers and due to the pandemic, our access and mobility was restricted. We expect our approach to work given the energy that such facilities are capable of generating. The trade-off that we will be seeking will be to avoid permanent damage to the DUT in order to obtain stable results and reliable evidence for law enforcement.

\subsection{Quantum Computers}
\label{sec:quantum}
Radioactive based SEUs maybe of great impact on quantum computers which have extreme sensitivity to minimal environmental variations e.g (temperature, vibrations) given their sub-atomic scale, even if shielded, not all sources of SEUs can be avoided~\cite{tibet, toronto}. In particular, we refer to the process of quantum decoherence and quantum error correction~\cite{qcorr} to obtain meaningful results from quantum computations. To this extent we consider of particular interest the NP problem of protein folding which can be greatly optimized by quantum annealing computers~\cite{folding} such as the one developed by D-Wave~\cite{dwave}, based on the quantum annealing meta-heuristic, fist proposed by Apolloni et al.~\cite{quantumA1} and furtherly developed by Kadowaki et al.~\cite{quantumA2}. In such process an error can lead to biological anomalies that can provoke diseases such as bovine spongiform encephalopathy~\cite{vacche}. It is trivial to evince how impactful this attack could be in a delicate processes such as vaccine development~\cite{rapidvax}, especially during pandemic times. Where making less errors can save human lives.

\subsection{Future works}
In future, we aim to explore less harmful, cheaper and more deterministic sources of SEUs i.e., lasers~\cite{laser,ddr}. Which, are capable of generating SEUs with high precision on classical machines, allowing our attack to be deployed outside a protected facility. Also, removing the epoxy layer of chips~\cite{cheapseu} requires very low effort to obtain SEUs with harmless radiations, though it damages the device irreversibly and it may raise concerns especially in forensics investigations. We will to test our approaches both on conventional and quantum machines.

\subsection{Countermeasures}
We have contacted the authors of Drammer and GuardION~\cite{drammer, guardion}, explaining our attack strategy. Our analysis lead to the conclusion that GuardION is not effective against externally provoked SEUs. Still, ECC (Error-correcting code) memories and magnetorestive memories (MRAM) are not susceptible to our kind of attack. Indeed they are used normally in high altitude operating devices. The impact of our approach resides on the fact that such memories are more expensive and slower than memory used in COTS devices. With respect to quantum computers, since their technology is very recent, several approaches could be taken to improve their safety and security, starting from the basic building blocks (Qbits) for example by realizing Qbit gates resistant to external interfences.


\section{Ethical Considerations}
Using ionizing radiations to provoke glitches in an electronic system is a potentially harmful, life threatening activity, which needs expert supervision and should never be deployed outside shielded environments. Generating such emissions maybe possible with relatively inexpensive material at home~\cite{ragazzoionico}, though such rudimentary techniques should never be pursued, not even for test purposes. This is the main reason why this paper is preliminary, because during pandemic times it was very hard to access safe facilities to complete extensive experiments.


\section{Conclusion}
Using radioactive soruce i.e. (Ionizing Radiations) to produce SEUs such as bit-flips is not a novel idea per-se~\cite{cazza, cazza1, cheapseu, clinical, ddr, toronto}. Our paper makes an effort to understand the possibility to use such events as a potential source of exploitation or a source of harmful errors on quantum machines. Our technique is helpful for law enforcement to unlock devices in a cheap and affordable way or to rescue a system, plus it can lead to safer quantum hardware development. Our preliminary results are promising and we are waiting to access a hardron-therapy facility to further verify our hypothesis.

\bibliographystyle{abbrv}
\bibliography{bib} 
\end{document}